\documentclass[twocolumn,showkeys,floatfix,aps,prd,9pt]{revtex4-2}
\usepackage{graphicx,epstopdf}
\usepackage[caption=false]{subfig}
\usepackage{appendix}
\usepackage[T1]{fontenc}
\usepackage{lmodern}
\usepackage[dvipsnames,x11names]{xcolor}
\usepackage[colorlinks=true,linkcolor=Magenta!80!black,citecolor=Magenta!70!black,urlcolor=Magenta!80!black]{hyperref}
\usepackage[sort&compress]{natbib}
\usepackage{morefloats}
\usepackage[pdf]{pstricks}
\usepackage{amsmath}
\usepackage{amssymb}
\usepackage{amsfonts}
\usepackage{rotating}
\usepackage{cancel}
\usepackage{mathtools}
\usepackage{bbm}
\usepackage{dsfont}
\usepackage{bbold}
\usepackage{multirow}
\usepackage{ulem}
\usepackage{physics}
\usepackage{orcidlink}
\usepackage{colortbl}
 \usepackage{booktabs}

\def\xslash{x\!\!\!\slash }

\begin{document}

\title{Flavor decomposition of electromagnetic multipole moments in $[q_1q_2][\bar q_3 \bar c]$ tetraquarks: Light-quark analogs of $T_{cs1}(2900)^0$ with $J^P = 1^+$}

\author{Ula\c{s}~\"{O}zdem\orcidlink{0000-0002-1907-2894}}%
\email[]{ulasozdem@aydin.edu.tr }
\affiliation{ Health Services Vocational School of Higher Education, Istanbul Aydin University, Sefakoy-Kucukcekmece, 34295 Istanbul, 
T\"{u}rkiye}

 
\begin{abstract}
We present a systematic study of the magnetic dipole and electric quadrupole moments of compact $[q_1 q_2][\bar q_3 \bar c]$ tetraquark states with quantum numbers $J^P = 1^+$, in the diquark--antidiquark picture and within the light-cone QCD sum rule framework. By 
performing a flavor decomposition across all admissible $u/d$ configurations, we show that the charm-quark contribution to the electromagnetic moments vanishes algebraically at the operator level -- a structural feature of the specific spin organization adopted here, in which the heavy antiquark resides in a pseudoscalar antidiquark, distinct from the dynamical $1/m_c$ suppression operative in alternative compact constructions and in hadronic molecules. The magnetic moments follow a coherent light-flavor pattern -- positive for $uu$-, negative for $dd$-, and suppressed for mixed-flavor diquarks -- and the quadrupole moments are non-zero throughout, corresponding, in the customary geometrical interpretation, to intrinsically deformed charge distributions. We further identify the dimensionless multipole ratio $\mathcal{D}/\mu$ as a structural observable, independent of the normalization of the interpolating current, which probes the relative geometry of the spin-current and charge distributions complementary to the absolute
moments. Taken together, the algebraic decoupling of the charm contribution and the operator-level distinction it implies among different singly-heavy tetraquark architectures constitute methodological signatures that discriminate among the corresponding internal organizations and, to the extent that a physical candidate is dominated by one of them, among compact and molecular interpretations.
\end{abstract}


\maketitle

\section{Motivation}\label{motivation}

The past two decades of hadron spectroscopy have produced a 
proliferation of states that defy the conventional quark-model
classification, ranging from the $X(3872)$~\cite{Belle:2003nnu} 
to a wide variety of $XYZ$ states, pentaquarks, and tetraquark 
candidates reported by LHCb, Belle, BESIII, and CMS. These 
discoveries have stimulated extensive theoretical efforts to 
unravel the internal structure of such states, as reviewed 
in~\cite{Esposito:2014rxa, Esposito:2016noz, Olsen:2017bmm, 
Lebed:2016hpi, Nielsen:2009uh, Brambilla:2019esw, Agaev:2020zad, 
Chen:2016qju, Ali:2017jda, Guo:2017jvc, Liu:2019zoy, Yang:2020atz, 
Dong:2021juy, Dong:2021bvy, Chen:2022asf, Meng:2022ozq, 
Wang:2025sic}.

A particularly active sector is the charm-strange one, where 
the LHCb collaboration has firmly established a family of 
tetraquark candidates with minimal quark content $\bar c s u\bar d$ 
clustered around $2.9~\text{GeV}$. These include the 
$T_{cs0}(2900)^0$ and $T_{cs1}(2900)^0$ with $J^P = 0^+$ and 
$1^-$~\cite{LHCb:2020pxc, LHCb:2020bls}, the $T_{a\,cs}(2900)$ 
isospin triplet with $I(J^P) = 1(0)^+$~\cite{LHCb:2022lzp, 
LHCb:2022sfr}, and the more recent confirmation of $T^{\ast}_{\bar c s}$ 
resonances in independent production channels~\cite{LHCb:2024vfz}. 
Theoretical analyses suggest that the same quark constituents 
should also bind into $J^P = 1^+$ and $2^+$ partners that have 
not yet been observed~\cite{Molina:2020hde, Dai:2022qwh, 
Sundu:2022kyd, Agaev:2021jsz, Mutuk:2020igv, Zhou:2025yjb, 
Zhou:2025rpb}. Discriminating among competing structural 
interpretations -- compact tetraquark, hadronic molecule, or 
other configurations -- requires observables beyond masses and 
decay widths, and electromagnetic multipole moments provide 
precisely such a diagnostic: the magnetic dipole moment $\mu$ 
probes the internal spin and current structure, while the 
electric quadrupole moment $\mathcal{D}$ measures the deviation 
from spherical symmetry of the charge distribution.

In this work we present the magnetic dipole and electric 
quadrupole moments of the $[q_1 q_2][\bar q_3 \bar c]$ 
($T_{q\bar c}$) tetraquark states with $J^P = 1^+$, taken as 
the light-quark analogs of the $T_{cs1}(2900)^0$ family with 
the strange quark replaced by a $u$ or $d$ quark, within the 
diquark--antidiquark picture. We work in the light-cone QCD 
sum rule (LCSR) framework in an external electromagnetic 
field~\cite{Chernyak:1990ag, Braun:1988qv, Balitsky:1989ry}, 
including photon distribution amplitudes up to twist-4 -- a 
framework that has been applied extensively to singly-heavy 
hadron systems~\cite{Agamaliev:2016wtt, Aliev:2015sea, 
Ozdem:2024ydl, Ozdem:2024pyb, Aliev:2008sk, Ozdem:2023edw, 
Ozdem:2023okg, Ozdem:2023eyz, Ozdem:2022ydv, Aliev:2015axa, 
Aliev:2014jra, Ozdem:2021vry, Azizi:2021aib, Aliev:2008ay, 
Azizi:2018jky, Aliev:2001ig, Azizi:2018mte, Ozdem:2025olj, 
Aliev:2001xr, Ozdem:2026wmf}.

The methodological contribution of this work goes beyond a direct 
application of the LCSR technique and rests on three observations that, 
taken together, define a clean structural signature of the present 
interpolating architecture. First, by performing a systematic flavor 
decomposition across all admissible light-quark configurations within 
the same diquark--antidiquark interpolation, we show that the 
charm-quark contribution to the electromagnetic moments vanishes 
algebraically rather than through a dynamical $1/m_c$ 
suppression: the charge $e_c$ drops out of the sum rules at the 
operator level, driven by the specific spin organization in which 
the charm antiquark resides in a pseudoscalar antidiquark. This 
algebraic decoupling is, to our knowledge, not articulated explicitly 
in previous LCSR analyses of singly heavy tetraquarks; we further 
show, by contrasting our result with diquark--antidiquark analyses 
in which the heavy quark instead sits in a vector antidiquark and 
with molecular configurations, that the algebraic vanishing is 
specific to the present spin organization, while alternative 
constructions yield non-vanishing charm contributions of either 
dynamical operator product expansion (OPE) or hadronic-molecule type. The character of the 
charm-quark contribution to the magnetic moment thus becomes an 
operator-level discriminator among inequivalent interpolations of 
the same quantum-numbered state. Second, we identify the dimensionless multipole ratio $\mathcal{D}/\mu$ at the photon point
as a structural observable that probes the relative weight of the quadrupole and magnetic
responses, independent of the overall normalization of the interpolating current and thereby
complementary to the absolute moments themselves. Third, the systematic flavor decomposition 
framework presented here provides a template that can be 
transferred to other singly heavy multiquark systems.

The paper is organized as follows. Section~\ref{formalism} 
introduces the theoretical framework, presents the interpolating 
current for the $T_{q\bar c}$ tetraquark, and derives the LCSR 
for the magnetic and quadrupole moments. Section~\ref{numerical} 
contains the numerical analysis. Section~\ref{sum} summarizes 
our findings.

 
\section{LCSR framework for electromagnetic multipole moments}\label{formalism}

The electromagnetic multipole moments of the  $T_{q\bar c}$ states ---specifically their magnetic dipole and electric quadrupole  moments---can be systematically extracted using the LCSR method. The natural starting point is the  correlation function
\begin{equation}
\label{corr}
T_{\mu\nu}(p,q) = i\int d^{4}x\, e^{ip\cdot x}\,
\langle 0|\mathcal{T}\{J_{\mu}(x)\, J_{\nu}^{\dagger}(0)\}|0\rangle_{F},
\end{equation}
where the subscript $F$ indicates that the vacuum expectation value is evaluated in the presence of the external electromagnetic field. 

In the LCSR approach, the interpolating current plays a central role, since it determines the quantum numbers and internal structure of the states that couple to the QCD vacuum. The current must therefore be constructed to have the same color, spin, and parity quantum numbers as the hadronic state under investigation.  
In the present work, the $T_{q\bar c}$ states are described by a tetraquark-type interpolating current composed of four quark fields,
\begin{align}
\label{curr1}
J_{\mu}(x) &=  \Big[q_{1a}^{T}(x) C \gamma_{\mu} q_{2b}(x)\Big]
\Big[\bar{q}_{3a}(x)\gamma_{5} C \bar{c}_{b}^{T}(x)\Big]  \nonumber\\
&- \Big[q_{1a}^{T}(x) C \gamma_{\mu} q_{2b}(x)\Big] 
\Big[\bar{q}_{3b}(x)\gamma_{5} C \bar{c}_{a}^{T}(x)\Big],
\end{align}
where $q_1$, $q_2$, and $q_3$ denote light $u$ or $d$ quarks, $C$ is the charge conjugation matrix, and $a$, $b$ are color indices.

The sum rules are obtained by matching two independent representations 
of $T_{\mu\nu}$. On the hadronic side, the correlator is 
saturated by intermediate states with the same quantum numbers as 
$J_\mu$; isolating the ground-state contribution yields a double-pole 
term proportional to the electromagnetic vertex 
$\langle T|J_\alpha^{em}|T\rangle$, which is decomposed into Lorentz 
structures associated with the magnetic dipole and electric 
quadrupole form factors. On the QCD side, the same correlator is 
evaluated through the operator product expansion near the light cone, 
with short-distance quark propagation treated perturbatively and 
long-distance photon-vacuum interactions encoded in photon 
distribution amplitudes of increasing twist~\cite{Ball:2002ps}. 
Equating the two representations, performing a double Borel 
transformation, and subtracting the continuum via quark-hadron 
duality yield the sum rules for the magnetic and quadrupole moments.

On the QCD side, the correlation function is evaluated by inserting the
interpolating currents into Eq.~\eqref{corr} and contracting the quark
fields using Wick's theorem. This procedure expresses the correlation
function in terms of quark propagators in the presence of the external
electromagnetic field. For the tetraquark currents considered in this work,
the resulting expression involves multiple Dirac and color traces. The full
expression can be written as
\begin{align}
T_{\mu\nu}^{\mathrm{QCD}}(p,q) &=- i \int d^4x\, e^{ip\cdot x} \langle 0 \mid
\Big\{
\mathrm{Tr}\big[\gamma_5 \tilde{S}_c^{b'b}(-x)\gamma_5 \nonumber\\
&\quad \times S_{q_3}^{a'a}(-x)\big]
\mathrm{Tr}\big[\gamma_\nu \tilde{S}_{q_1}^{aa'}(x)\gamma_\mu S_{q_2}^{bb'}(x)\big]
\nonumber\\
&\quad- \mathrm{Tr}\big[\gamma_5 \tilde{S}_c^{a'b}(-x)\gamma_5 S_{q_3}^{b'a}(-x)\big]\nonumber\\
&\quad
\times \mathrm{Tr}\big[\gamma_\nu \tilde{S}_{q_1}^{aa'}(x)\gamma_\mu S_{q_2}^{bb'}(x)\big]
\nonumber\\
&\quad- \mathrm{Tr}\big[\gamma_5 \tilde{S}_c^{b'a}(-x)\gamma_5 S_{q_3}^{a'b}(-x)\big]
\nonumber\\
&\quad \times
\mathrm{Tr}\big[\gamma_\nu \tilde{S}_{q_1}^{aa'}(x)\gamma_\mu S_{q_2}^{bb'}(x)\big] \nonumber\\
&\quad
+ \mathrm{Tr}\big[\gamma_5 \tilde{S}_c^{a'a}(-x)\gamma_5 S_{q_3}^{b'b}(-x)\big]
\nonumber\\
&\quad \times
\mathrm{Tr}\big[\gamma_\nu \tilde{S}_{q_1}^{aa'}(x)\gamma_\mu S_{q_2}^{bb'}(x)\big]
\Big\}\mid 0 \rangle_{F},
\label{eq:QCDfull}
\end{align}
where $ \tilde{S}_q^{ab}(x) = C S_q^{T\,ab}(x) C$ denotes the charge-conjugated quark propagator, and the color indices
$a, a',b, b'$ are summed implicitly.  The overall expression is evaluated in
the presence of the external electromagnetic field, indicated by the
subscript $F$.

The calculation is performed in coordinate space using the light- and
heavy-quark propagators in the presence of background gluon fields.
We work in the fixed-point gauge $x^\mu A_\mu(x)=0$ (Fock-Schwinger gauge), which simplifies the expansion of the propagators in terms of the gluon field strength. For a light quark ($q=u,d$), the propagator is given
by~\cite{Balitsky:1987bk}
\begin{align}
S_{q}(x) &=\frac{i\xslash}{2\pi^{2}x^{4}} - \frac{m_{q}}{4\pi^{2}x^{2}} 
- \frac{i g_{s}}{16\pi^{2}x^{2}} \int_{0}^{1} du\; G^{\mu\nu}(ux)\nonumber\\
& \times 
\Big[ \bar u \xslash\sigma_{\mu\nu} + u \sigma_{\mu\nu} \xslash\Big],
\label{edmn13}
\end{align}
while for the charm quark one has~\cite{Belyaev:1985wza}

\begin{align}
S_{c}(x) &= \frac{m_{c}^{2}}{4\pi^{2}}\Big[ 
\frac{K_{1}\!\big(m_{c}\sqrt{-x^{2}}\big)}{\sqrt{-x^{2}}}
+ i\frac{\xslash K_{2}\!\big(m_{c}\sqrt{-x^{2}}\big)}
{(\sqrt{-x^{2}})^{2}} \Big] 
\nonumber\\
&- \frac{i m_{c} g_{s}}{16\pi^{2}} \int_{0}^{1} du\; G^{\mu\nu}(ux)
\Big[ (\sigma_{\mu\nu}\xslash+ \xslash \sigma_{\mu\nu})
 \nonumber
\end{align}
\begin{align}
& \times \frac{K_{1}\!\big(m_{c}\sqrt{-x^{2}}\big)}{\sqrt{-x^{2}}} 
+ 2\sigma_{\mu\nu} K_{0}\!\big(m_{c}\sqrt{-x^{2}}\big) \Big].
\label{edmn14}
\end{align}
where $G^{\mu\nu}$ is the gluon field-strength tensor, $K_n$ are modified Bessel functions of the second kind, and we set $m_{u,d}=0$ throughout. The leading term of both the light- and heavy-quark propagators represents the free part, whereas the subsequent terms encode the non-perturbative interactions with the gluon background. 

The interaction of the photon with the quark system occurs through both perturbative and non-perturbative mechanisms, which are treated separately.

\noindent
Perturbative photon emission: 
In the short-distance regime, the photon couples directly to a quark line
through the QED vertex. This contribution is obtained by replacing one of
the quark propagators in Eq.~\eqref{eq:QCDfull} according to
\begin{align}
S_{q(c)}^{\mathrm{free}}(x)
\;\rightarrow\;
\int d^{4}z\;
S_{q(c)}^{\mathrm{free}}(x-z)\,
\gamma^\mu A_\mu(z)\,
S_{q(c)}^{\mathrm{free}}(z),
\label{free}
\end{align}
while the remaining propagators are kept in their free form. 
Here, $A_{\mu}(z)$ denotes the electromagnetic potential 
corresponding to the external photon field of momentum $q$ 
and polarization $\varepsilon$, treated as a plane wave 
in the linear-response approximation. This term encodes the 
short-distance contribution in which the photon couples 
directly to a quark line via the QED vertex, in contrast to 
the non-perturbative coupling discussed below.

\noindent
Non-perturbative photon emission: 
At long distances, the photon interacts with the QCD vacuum through
light-quark condensates and photon distribution amplitudes (DAs).
This effect is incorporated by replacing one of the light-quark propagators
in Eq.~\eqref{eq:QCDfull} with its vacuum expectation value in the presence
of the external photon field. Schematically,
\begin{align}
S_{q,\alpha\beta}^{ab}(x)
\;\rightarrow\;
-\frac{1}{4}
\left[
\bar{q}^{a}(x)\Gamma_{i}q^{b}(0)
\right]
(\Gamma_{i})_{\alpha\beta},
\label{edmn21}
\end{align}
where the sum runs over the Dirac structures
$ 
\Gamma_{i} =
\{
\mathbb{1},
\gamma_{5},
\gamma_{\mu},
i\gamma_{5}\gamma_{\mu},
\sigma_{\mu\nu}/2
\}$.

The resulting nonlocal matrix elements $\langle \gamma(q) | \bar{q}(x)\Gamma_{i}q(0) |0\rangle$, $\langle \gamma(q) |
\bar{q}(x)\Gamma_{i}G_{\alpha\beta}q(0) |0\rangle$, are expressed in terms of photon DAs of increasing twist~\cite{Ball:2002ps}. In the present analysis, non-perturbative photon couplings are included only for light quarks ($u,d$). Contributions involving heavy-quark photon DAs are neglected since they are suppressed by powers of the heavy-quark mass $1/m_c$, in line with the negligible charm content of the QCD vacuum~\cite{Antonov:2012ud}. Consequently, heavy quarks contribute only through perturbative photon emission, while the non-perturbative structure of the photon is entirely encoded in the light-quark sector.

Combining all perturbative and non-perturbative contributions in
Eq.~\eqref{eq:QCDfull} yields the complete QCD representation of the
correlation function.

The hadronic representation of the correlation function is obtained by
inserting a complete set of hadronic states with the same quantum numbers as the interpolating current into Eq.~\eqref{corr}. Isolating the contribution of the lowest-lying tetraquark state $T_{q\bar c}$ with quantum numbers $J^P=1^+$, the correlation function can be written as
\begin{align}
\label{edmn04}
T_{\mu\nu}^{\mathrm{Had}}(p,q) &=
\frac{
\langle 0 | J_\mu | T_{q\bar c}(p,\varepsilon^i) \rangle
\langle T_{q\bar c}(p,\varepsilon^i) |
T_{q\bar c}(p',\varepsilon^f) \rangle_F
}{
(p^2 - m_{T_{q\bar c}}^2)(p'^2 - m_{T_{q\bar c}}^2)
}\nonumber\\
&
  \times \langle T_{q\bar c}(p',\varepsilon^f) | J_\nu^\dagger | 0 \rangle + \cdots ,
\end{align}
where $p' = p + q$, and the ellipsis denotes contributions from higher
resonances and continuum states.

The matrix elements describing the couplings of the interpolating current to the initial and final physical tetraquark states are parametrized as  $
\langle 0 | J_\mu | T_{q\bar c}(p,\varepsilon^{i}) \rangle
= \lambda_{T_{q\bar c}} \, \varepsilon_\mu^{i}$ and 
$ \langle T_{q\bar c}(p',\varepsilon^{f}) | J_\nu^\dagger | 0 \rangle
= \lambda_{T_{q\bar c}} \, \varepsilon_\nu^{f\,*}$,
where $\lambda_{T_{q\bar c}}$ is the coupling constant (pole residue),  and $\varepsilon_\mu^{i}$ and $\varepsilon_\nu^{f}$ denote the polarization vectors of the initial and final axial-vector 
tetraquark states, respectively. These tetraquark polarization 
vectors should be distinguished from the photon polarization 
vector $\varepsilon^{\gamma}$, which describes the external 
electromagnetic field and will appear explicitly in 
Eq.~\eqref{edmn06} below.

The interaction of the tetraquark with the external electromagnetic field is described by the matrix element 
$\langle T_{q\bar c}(p,\varepsilon^i) |
T_{q\bar c}(p',\varepsilon^f) \rangle_F ,$
which can be decomposed in terms of three independent Lorentz-invariant
form factors $G_1(Q^2)$, $G_2(Q^2)$, and $G_3(Q^2)$  as follows~\cite{Arnold:1979cg, Brodsky:1992px}:
\begin{align}
\label{edmn06}
\langle T_{q\bar c}(p,\varepsilon^i) |
T_{q\bar c}(p',\varepsilon^f) \rangle_F
& = - \varepsilon^\gamma (\varepsilon^{i})^\mu (\varepsilon^{f})^\nu \nonumber\\
& \times 
\Big[
G_1(Q^2) (p+p')_\gamma g_{\mu\nu}
\nonumber\\
&
+ G_2(Q^2) \big( g_{\gamma\nu} q_\mu - g_{\gamma\mu} q_\nu \big)
\nonumber\\
&
- \frac{G_3(Q^2)}{2m_{T_{q\bar c}}^2} (p+p')_\gamma q_\mu q_\nu
\Big],
\end{align}
where $\varepsilon^\gamma$ is the photon polarization vector and
$Q^2 = -q^2$ is the momentum transfer squared.

It is convenient to introduce the magnetic and quadrupole form factors, $F_M(Q^2)$ and $F_{\mathcal{D}}(Q^2)$, which are related to the invariant
form factors by
\begin{align}
F_M(Q^2) &= G_2(Q^2),
\\
F_{\mathcal{D}}(Q^2) &=
G_1(Q^2)
-
G_2(Q^2)
+
\Big(
1 + \frac{Q^2}{4m_{T_{q\bar c}}^2}
\Big)
G_3(Q^2).\label{edmn07} 
\end{align}

In the static limit corresponding to real photons ($Q^2 = 0$),
these form factors determine the electromagnetic multipole moments:
\begin{align}
\mu_{T_{q\bar c}} &=
\frac{e}{2m_{T_{q\bar c}}}
F_M(0),
\qquad 
\mathcal{D}_{T_{q\bar c}} =
\frac{e}{m_{T_{q\bar c}}^2}
F_{\mathcal{D}}(0).\label{edmn0008}
\end{align}

Substituting Eqs.~\eqref{edmn06}--\eqref{edmn07} into
Eq.~\eqref{edmn04}, one obtains the hadronic representation
of the correlation function. Keeping only the structures relevant
for the extraction of the magnetic and quadrupole moments yields
 %
\begin{align}
\label{finalhad}
T^{\mathrm{Had}}_{\mu\nu}(p,q) &=
\frac{\lambda_{T_{q\bar c}}^{2}}
{
(m_{T_{q\bar c}}^2 - p^2)
(m_{T_{q\bar c}}^2 - p'^2)
}
\Big[
F_M(0)\,
\varepsilon^{\gamma}
\big(
g_{\gamma\nu}\, q_\mu \nonumber\\
&
-
g_{\gamma\mu}\, q_\nu
\big)
-
F_{\mathcal{D}}(0)
\frac{\varepsilon^{\gamma} p_{\gamma}}{m_{T_{q\bar c}}^2}\,
q_\mu q_\nu
+ \cdots
\Big],
\end{align}
where the ellipsis denotes additional Lorentz structures that do not contribute to
the sum rules used to extract the electromagnetic moments.  Here $\varepsilon^{\gamma}$ is the photon polarization vector of Eq.~(\ref{edmn06}), the
polarizations of the tetraquark states having been summed over using
$\sum \varepsilon_\mu \varepsilon^{*}_\nu = -g_{\mu\nu} + p_\mu p_\nu/m_{T_{q\bar c}}^2$. 
This expression constitutes the hadronic side of the sum rules. Matching
it to the QCD representation after performing the Borel transformation
allows one to determine the magnetic and quadrupole moments of the
$T_{q\bar c}$ tetraquark.

Finally, after performing the spectral integrals and equating the hadronic
and QCD representations, we obtain the LCSR for the
electromagnetic moments:
\begin{align}
\mu_{T_{q\bar c}} \,\lambda_{T_{q\bar c}}^2
&=
e^{m_{T_{q\bar c}}^2/M^2}
\,
\mathcal{R}_\mu(M^2,s_0),
\label{jmu1}
\\
\mathcal{D}_{T_{q\bar c}} \, \lambda_{T_{q\bar c}}^2
&=
e^{m_{T_{q\bar c}}^2/M^2}
\,
\mathcal{R}_{\mathcal{D}}(M^2,s_0).
\label{jmu3}
\end{align}

The functions $\mathcal{R}_\mu(M^2,s_0)$ and $\mathcal{R}_{\mathcal{D}}(M^2,s_0)$
represent the QCD spectral integrals corresponding to the magnetic and quadrupole structures, respectively. 
Their explicit expressions are  presented in Appendix.


\section{Numerical Aspects}\label{numerical}

The numerical analysis of the magnetic dipole and quadrupole moments within the LCSR framework requires a set of input parameters that encode the nonperturbative structure of the QCD vacuum. In this work, all input parameters are taken from widely used values available in the literature. These include the charm quark mass $m_c = 1.2730 \pm 0.0046~\mathrm{GeV}$~\cite{ParticleDataGroup:2024cfk}; the mass $m_{T_{q\bar c}} = 2.843^{+0.134}_{-0.157}~\mathrm{GeV}$ and residue $\lambda_{T_{q\bar c}} = (7.60 \pm 1.80) \times 10^{-3}~\mathrm{GeV}^5$ of the tetraquark states~\cite{Agaev:2017oay}; the light-quark condensate $\langle \bar q q \rangle = (-0.24 \pm 0.01)^3~\mathrm{GeV}^3$~\cite{Ioffe:2005ym}; the gluon condensate $\langle g_s^2 G^2 \rangle = 0.48 \pm 0.14~\mathrm{GeV}^4$~\cite{Narison:2018nbv}; the photon DA constant $f_{3\gamma} = -0.0039~\mathrm{GeV}^2$~\cite{Ball:2002ps}; and the magnetic susceptibility $\chi = 2.85 \pm 0.5~\mathrm{GeV}^{-2}$ of the quark condensate~\cite{Rohrwild:2007yt}. The up- and down-quark masses are neglected by setting $m_u = m_d = 0$. The photon DAs are taken from~\cite{Ball:2002ps}, and contributions up to twist-4 accuracy are included in the calculation.

According to Eqs.~(\ref{jmu1})--(\ref{jmu3}), the LCSR predictions depend on two auxiliary parameters: the continuum threshold $s_0$ and the Borel mass parameter $M^2$. Although these parameters are not physical observables, reliable determinations of the magnetic and quadrupole moments require identifying regions in which the results exhibit weak dependence on their variations. These regions define the working windows of the sum rules. 
The continuum threshold $s_0$ separates the ground-state contribution from the excited states and continuum in the hadronic representation of the correlation function. Phenomenologically, $s_0$ is expected to lie slightly above the squared mass of the lowest-lying state. Following standard QCD sum-rule practice, we adopt the interval $\sim$
$ (m_H + 0.6)^2~\mathrm{GeV}^2 \leq s_0 \leq (m_H + 0.75)^2~\mathrm{GeV}^2 ,$ 
where $m_H$ denotes the mass of the hadronic state under consideration. This choice is known to yield stable predictions in similar analyses and is used throughout this work. 
The working region of the Borel parameter $M^2$ is determined by two competing requirements. The lower bound is fixed by the convergence of the OPE, requiring that higher-dimensional condensate contributions remain suppressed. The upper bound is constrained by pole dominance, ensuring that the ground-state contribution is not overwhelmed by the continuum. These conditions are quantified through the pole contribution (PC) and convergence (CVG) criteria:
\begin{align}
\mathrm{PC} &=
\frac{\mathcal{R}_{\mu(\mathcal D)}(M^2,s_0)}
     {\mathcal{R}_{\mu(\mathcal D)}(M^2,\infty)}
> 40\% , \\
\mathrm{CVG} &=
\frac{\mathcal{R}_{\mu(\mathcal D)}^{\mathrm{Dim\,7}}(M^2,s_0)}
     {\mathcal{R}_{\mu(\mathcal D)}(M^2,s_0)}
< 5\% ,
\end{align}
where $\mathcal{R}_{\mu(\mathcal D)}(M^2,s_0)$ denotes the invariant function entering the sum rules, and $\mathcal{R}_{\mu(\mathcal D)}^{\mathrm{Dim\,7}}(M^2,s_0)$ represents the highest-dimensional contribution retained in the OPE.
\begin{widetext}

\begin{table}[htb]
\caption{
Working windows of the LCSR auxiliary parameters $s_0$ and $M^2$ 
for the electromagnetic observables of the 
${[q_1 q_2][\bar q_3 \bar c]}$ systems. 
The corresponding ranges of the PC and 
the relative dimension-7 contribution are also shown.}
\label{table}
\begin{ruledtabular}
\begin{tabular}{l|ccccc}

Moment  &Spectral Density & $s_0\,(\mathrm{GeV}^2)$ & $M^2\,(\mathrm{GeV}^2)$ & PC $(\%)$ & CVG $(\%)$ 
\\
\hline

$\mu$ & $\mathcal{R}_\mu(M^2,s_0)$ & $12.0 - 13.0$ & $1.5 - 1.8$ 
& $41.9 - 58.3$ & $ 0.46$ 
\\
$\mathcal D$ & $\mathcal{R}_{\mathcal D}(M^2,s_0)$ & $12.0 - 13.0$ 
& $1.5 - 1.8$  & $40.6 - 58.0$  & $ 0.10$ 
\\
\end{tabular}
\end{ruledtabular}
\end{table}

\end{widetext}

\begin{figure}[htb!]
\includegraphics[width=0.4\textwidth]{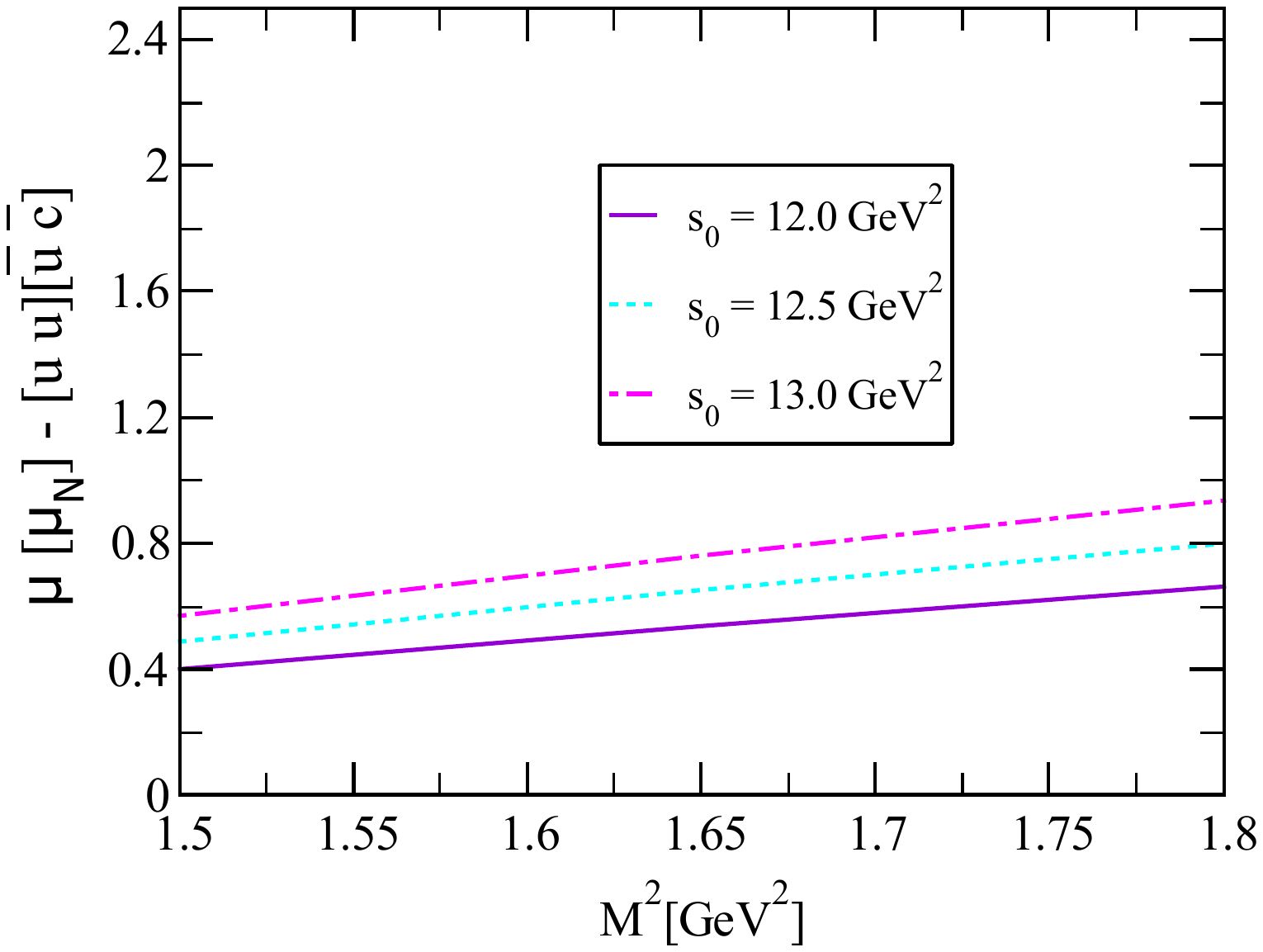}
\includegraphics[width=0.4\textwidth]{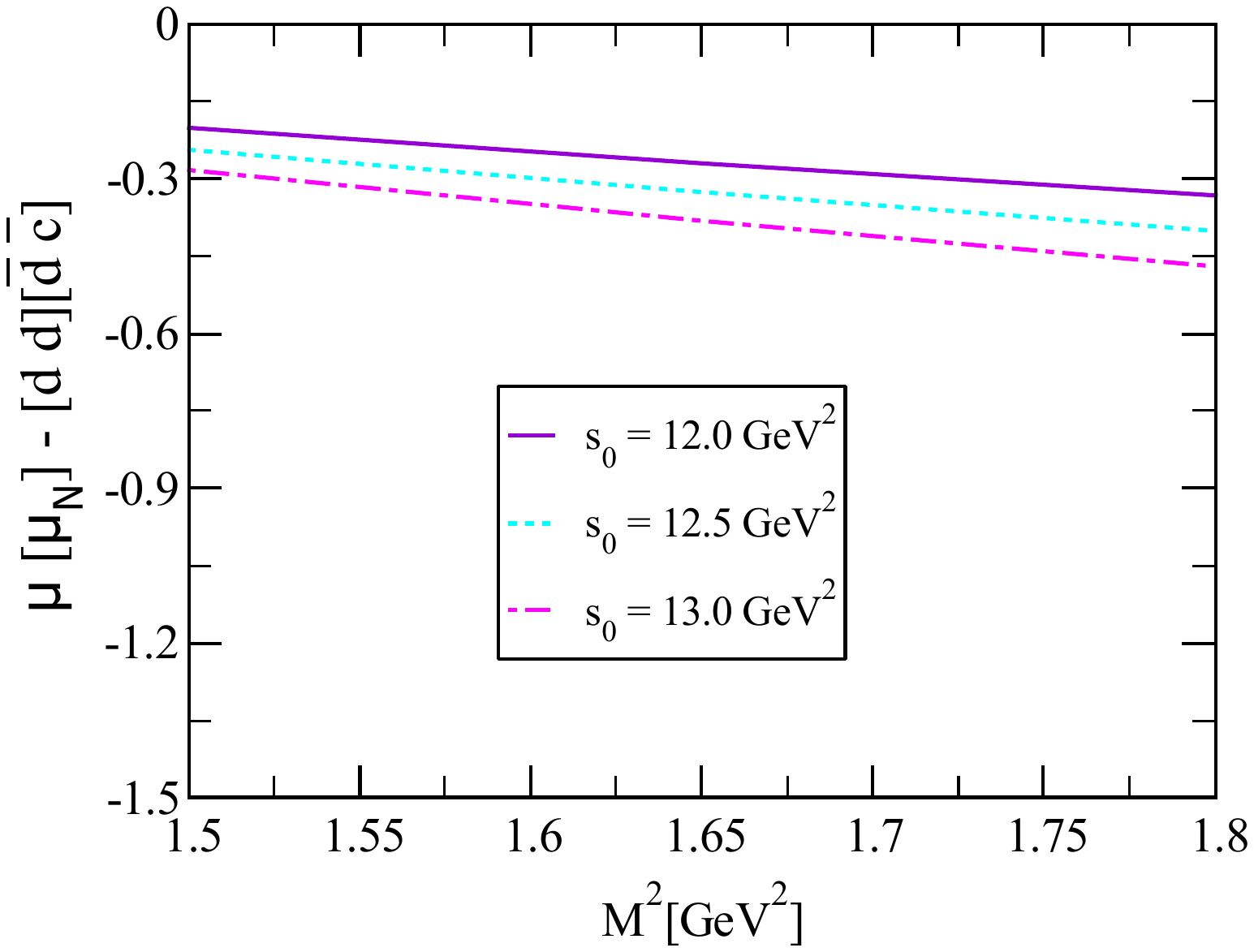}
\caption{Dependence of the extracted magnetic dipole moment on the Borel mass $M^2$ for the $[uu][\bar u \bar c]$ and $[dd][\bar d \bar c]$ configurations.}
\label{Msqfig}
\end{figure}

By imposing these criteria simultaneously, we determine the working 
regions for $M^2$ and $s_0$ summarized in Table~\ref{table}. Within 
these windows, the pole contribution varies between $40\%$ and 
$60\%$, confirming the dominance of the lowest-lying state and 
justifying the single-pole approximation employed in the hadronic 
parametrization. At the same time, the OPE exhibits good 
convergence: the relative contribution of the dimension-7 terms 
remains below $1\%$ over the entire Borel window, well within the 
adopted $5\%$ criterion.  
The simultaneous fulfillment of the PC and CVG requirements defines the fiducial Borel working region in which the sum rules are reliable. 
As an illustrative example, Fig.~\ref{Msqfig} shows the 
resulting magnetic moment as functions of the Borel parameter $M^2$ 
for several fixed values of $s_0$ in the ${[u u][\bar u \bar c]}$ 
and ${[dd][\bar d \bar c]}$ channels. One observes that the extracted magnetic moment depends moderately on the auxiliary
parameters $M^2$ and $s_0$, a sensitivity that originates in the perturbative contribution
and is discussed further in connection with the multipole ratio below. These variations are
included in the quoted theoretical uncertainties.

A comprehensive numerical analysis is performed to extract the magnetic and electric quadrupole moments of the ${[q_1 q_2][\bar q_3 \bar c]}$ tetraquark states. The resulting central values and uncertainties are summarized in Table~\ref{table1}. The quoted errors include the systematic propagation of all input parameters as well as the variations of the auxiliary LCSR parameters, $M^2$ and $s_0$. The relative contributions to the total uncertainty are estimated to be approximately $15\%$ from the tetraquark masses, $10\%$ from quark masses, $20\%$ from the residue parameters, $30\%$ from the continuum threshold $s_0$, $10\%$ from the Borel mass parameter $M^2$, $10\%$ from the photon DAs, and about $5\%$ from remaining sources such as QCD condensates. All individual uncertainties are combined in quadrature to obtain the final error bars reported in Table~\ref{table1}.
\begin{table}[htb!]
\caption{Magnetic dipole (in units of $\mu_N$) and electric quadrupole (in units of $\times 10^{-2}\,\mathrm{fm}^2$) moments of the ${[q_1 q_2][\bar q_3 \bar c]}$ compact tetraquark states.}
\label{table1}
\begin{ruledtabular}
\begin{tabular}{l|cccccccc}

 Moment  & $[uu][\bar u \bar c]$ & $[uu][\bar d \bar c]$ &
$[ud][\bar d \bar c]$ & $[ud][\bar u \bar c]$ \\

\hline\hline

 $\mu  $  & $0.66 \pm 0.23$ & $0.87 \pm 0.30$ & $0.27 \pm 0.10$
& $0.06 \pm 0.02$ \\
 $\mathcal{D}$ & $1.99 \pm 0.20$ & $1.97 \pm 0.20$ & $0.49 \pm 0.05$
& $0.49 \pm 0.05$ \\

\hline\hline

 Moment & $[dd][\bar d \bar c]$ & $[dd][\bar u \bar c]$ &
$[du][\bar u \bar c]$ & $[du][\bar d \bar c]$ \\

\hline\hline

 $\mu $ 
& $-0.33 \pm 0.11$ & $-0.54 \pm 0.19$ & $0.06 \pm 0.02$
& $0.27 \pm 0.10$ \\ 

 $\mathcal{D}$  & $-0.99 \pm 0.10$ & $-0.97 \pm 0.10$
& $0.49 \pm 0.05$ & $0.49 \pm 0.05$ \\

\end{tabular}
\end{ruledtabular}
\end{table}

Our analysis provides quantitative information on the electromagnetic structure of the 
${[q_1 q_2][\bar q_3 \bar c]}$ compact tetraquark configurations. The numerical results
presented in Tables~\ref{table1}, \ref{table2} and \ref{table3} allow us to identify several systematic
patterns in the magnetic and quadrupole moments, which reflect the internal quark
composition and spatial distributions of the states.

The magnetic moments listed in Table~\ref{table1} display a coherent 
flavor pattern: states whose light diquark consists of two $u$ quarks 
produce the largest positive magnetic moments ($0.66 - 0.87\,\mu_N$), 
those with a $dd$ diquark give negative values ($-0.33$ to 
$-0.54\,\mu_N$), and configurations with a mixed $ud$ diquark yield 
significantly smaller magnitudes ($0.06 - 0.27\,\mu_N$) through 
partial cancellations between the individual quark contributions. 
A finer structure within each class is set by the light antiquark 
$\bar q_3$: comparing $[uu][\bar u\bar c]$ ($\mu = 0.66\,\mu_N$) with 
$[uu][\bar d\bar c]$ ($\mu = 0.87\,\mu_N$), the difference of 
$0.21\,\mu_N$ traces directly to the opposite signs of the 
antiquark contributions, $\mu_{\bar u} = -0.14\,\mu_N$ versus 
$\mu_{\bar d} = 0.07\,\mu_N$ (Table~\ref{table2}), reflecting the 
electric-charge structure of the antiquark sector. The same 
$\bar u \leftrightarrow \bar d$ replacement induces a comparable 
shift in every $dd$-diquark state, indicating that the antiquark 
flavor acts as a tunable handle on top of the dominant diquark 
contribution.

Two structural features of these results deserve particular emphasis, 
as they reflect the methodological character of the calculation rather 
than its numerical output.

First, the flavor decomposition of Table~\ref{table2} shows that the
charm-quark contribution to every magnetic moment vanishes,
$\mu_c = 0$, across all eight flavor configurations. This is neither
a numerical artifact nor a dynamical $1/m_c$ suppression: it is an
algebraic consequence of the operator structure of the interpolating
current of Eq.~(\ref{curr1}). When the electromagnetic vertex is
inserted perturbatively on the charm line of the correlation
function, the Dirac trace associated with the pseudoscalar
antidiquark, in which the charm antiquark couples only through
$\gamma_5$, eliminates all terms proportional to $e_c$ from the
analytic expressions exactly. The charm mass $m_c$ continues to set
the threshold and the spectral structure of the calculation, but
carries no electromagnetic weight. The remaining charm--photon
couplings would proceed through nonperturbative charm matrix
elements -- the heavy-quark photon DAs and the
charm condensate -- which are power suppressed in $1/m_c$,
reflecting the negligible charm content of the QCD vacuum, and are
consistently neglected in the present framework
(Sec.~\ref{formalism}). Within this framework, $\mu_c = 0$
therefore holds identically, with any residual charm sensitivity
confined to power-suppressed terms. The total moments are thus
determined entirely by the light-quark contributions $\mu_{q_1}$,
$\mu_{q_2}$, $\mu_{q_3}$, as confirmed explicitly in
Table~\ref{table2}. To our knowledge, this algebraic decoupling has
not been articulated explicitly in previous LCSR analyses of singly
heavy tetraquarks, where heavy-quark contributions are usually
treated as dynamically suppressed and quoted only as a numerical
smallness.

This decoupling is a property of the interpolating architecture
rather than of the physical state. A general interpolating field
with $J^P = 1^+$ and the quark content considered here is a linear
combination of inequivalent diquark--antidiquark structures -- most
notably the present one, with an axial-vector light diquark and a
pseudoscalar heavy antidiquark, and the inverted construction in
which the Lorentz index resides on the heavy antidiquark, as
employed for $X_{\mathrm{AV}}$ in Ref.~\cite{Ozdem:2023edw} -- and a
physical state is in general expected to couple to all of them. The
algebraic decoupling holds at the pure axial-light point of this
operator space: for the inverted construction the charm contribution
is non-vanishing and comparable to the light-quark
terms~\cite{Ozdem:2023edw}, and for a generic admixture the charge
$e_c$ re-enters the sum rules with a strength controlled by the
admixture. The prediction $\mu_c = 0$ therefore applies to a
physical state to the extent that its internal organization is
dominated by the present configuration, and, conversely, the
charm-sensitivity pattern of the moments provides an operator-level
handle on which configuration dominates. Since the flavor
decomposition itself is not experimentally accessible, this
diagnostic operates in practice through the distinct total moments
predicted by the inequivalent architectures, as quantified in the
comparison below. 

Second, the calculation contains a non-trivial internal consistency 
check. The interpolating current of Eq.~(\ref{curr1}) is symmetric 
under the exchange $q_1 \leftrightarrow q_2$ of the two light quarks 
inside the axial-vector diquark, since the combination 
$q_1^T C\gamma_\mu q_2$ is symmetric in the flavor labels for the 
$J^P = 1^+$ current. Configurations related by this exchange must 
therefore yield identical magnetic moments. This is confirmed in 
the data: $[ud][\bar u\bar c]$ and $[du][\bar u\bar c]$ both give 
$\mu = 0.06\,\mu_N$, and $[ud][\bar d\bar c]$ and $[du][\bar d\bar c]$ 
both give $\mu = 0.27\,\mu_N$. The eight entries of Table~\ref{table1} 
therefore reduce to six physically distinct values, with two pairs 
of equivalent configurations providing a non-trivial check of the 
sum-rule calculation. In addition, the magnitudes of states related 
by the global $u \leftrightarrow d$ interchange are comparable up 
to a sign (e.g., $[uu][\bar d\bar c]$ at $0.87\,\mu_N$ versus 
$[dd][\bar u\bar c]$ at $-0.54\,\mu_N$), with the small residual 
difference reflecting the asymmetric position of the antiquark in 
the antidiquark slot.

\begin{table}[htb!]
\caption{
Individual quark-flavor contributions to the magnetic dipole moments 
of the compact ${[q_1 q_2][\bar q_3 \bar c]}$ tetraquark states with 
quantum numbers $J^P = 1^+$. All quantities are given in units of the nuclear magneton $\mu_N$. The columns $\mu_{q_1}$, $\mu_{q_2}$, and $\mu_{q_3}$ denote the contributions of the light quarks and antiquarks, while $\mu_c$ 
corresponds to the charm-quark contribution. }
\label{table2}
\addtolength{\tabcolsep}{6pt}
\begin{ruledtabular}
\begin{tabular}{l|ccccc}

States & $\mu_{q_1}$ & $\mu_{q_2}$ & $\mu_{q_3}$ & $\mu_{c}$ & $\mu_{\mathrm{tot}}$ \\

\hline\hline
$[uu][\bar u \bar c]$ & $~~0.40$ & $~~0.40$ & $-0.14$ & $0$ & $~~0.66$ \\
$[uu][\bar d \bar c]$ & $~~0.40$ & $~~0.40$ & $~~0.07$  & $0$ & $~~0.87$ \\
$[ud][\bar d \bar c]$ & $~~0.40$ & $-0.20$ & $~~0.07$ & $0$ & $~~0.27$ \\
$[ud][\bar u \bar c]$ & $~~0.40$ & $-0.20$ & $-0.14$ & $0$ & $~~0.06$ \\
$[dd][\bar d \bar c]$ & $-0.20$ & $-0.20$ & $~~0.07$ & $0$ & $-0.33$ \\
$[dd][\bar u \bar c]$ & $-0.20$ & $-0.20$ & $-0.14$ & $0$ & $-0.54$ \\
$[du][\bar u \bar c]$ & $-0.20$ & $~~0.40$ & $-0.14$ & $0$ & $~~0.06$ \\
$[du][\bar d \bar c]$ & $-0.20$ & $~~0.40$ & $~~0.07$ & $0$ & $~~0.27$ \\
\end{tabular}
\end{ruledtabular}
\end{table}

The electric quadrupole moments presented in Table~\ref{table1}
provide complementary information on the charge structure of the
tetraquark states, and are found to be nonzero in all channels
considered here. We first state the precise meaning of the
geometrical language employed below. What the sum rules determine is
the quadrupole form factor at the photon point,
$F_{\mathcal D}(0)$, a property of the electromagnetic vertex; its
translation into a three-dimensional charge density is
representation dependent for a relativistic bound state, and the
notions of deformation and of prolate or oblate shapes are therefore
used here in the conventional, schematic Breit-frame sense familiar
from the deuteron and the $\Delta(1232)$. Within this convention the
statement is structural: a purely $s$-wave configuration with an
isotropic charge organization would yield $\mathcal{D} = 0$, so the
systematically non-vanishing values obtained here reflect an
intrinsic quadrupole coupling of the diquark--antidiquark system
rather than an accidental feature of any particular flavor channel.

\begin{table}[htb!]
\addtolength{\tabcolsep}{6pt}
\caption{ Individual quark-flavor contributions to the electric quadrupole moments of the compact ${[q_1 q_2][\bar q_3 \bar c]}$ tetraquark states with 
quantum numbers $J^P = 1^+$. All quantities are given in units of $\times 10^{-2}\,\mathrm{fm}^2$. The columns $\mathcal{D}_{q_1}$, $\mathcal{D}_{q_2}$, and 
$\mathcal{D}_{q_3}$ denote the contributions of the light quarks 
and antiquarks, while $\mathcal{D}_c$ corresponds to the charm-quark 
contribution.}
\label{table3}
\begin{ruledtabular}
\begin{tabular}{l|ccccc}

States & $\mathcal D_{q_1}$ & $\mathcal D_{q_2}$ & $\mathcal D_{q_3}$ & $\mathcal D_{c}$ & $\mathcal D_{\mathrm{tot}}$ \\

\hline\hline
$[uu][\bar u \bar c]$ & $~~0.99$ & $~~0.99$ & $~~0.01$ & $0$ & $~~1.99$ \\
$[uu][\bar d \bar c]$ & $~~0.99$ & $~~0.99$ & $-0.01$ & $0$ & $~~1.97$ \\
$[ud][\bar d \bar c]$ & $~~0.99$ & $-0.49$ & $-0.01$ & $0$ & $~~0.49$ \\
$[ud][\bar u \bar c]$ & $~~0.99$ & $-0.49$ & $~~0.01$ & $0$ & $~~0.49$ \\
$[dd][\bar d \bar c]$ & $-0.49$ & $-0.49$ & $-0.01$ & $0$ & $-0.99$ \\
$[dd][\bar u \bar c]$ & $-0.49$ & $-0.49$ & $~~0.01$ & $0$ & $-0.97$ \\
$[du][\bar u \bar c]$ & $-0.49$ & $~~0.99$ & $~~0.01$ & $0$ & $~~0.49$ \\
$[du][\bar d \bar c]$ & $-0.49$ & $~~0.99$ & $-0.01$ & $0$ & $~~0.49$ \\
\end{tabular}
\end{ruledtabular}
\end{table}

In this interpretation, the sign of $\mathcal{D}$ encodes the character of the deformation: 
negative values correspond to oblate charge distributions (flattened 
along the symmetry axis), while positive values indicate prolate 
configurations (elongated along the symmetry axis). The pattern 
emerging from Table~\ref{table1} is striking. States with two $u$ 
quarks yield positive quadrupole moments around $2.0 \times 
10^{-2}~\mathrm{fm}^2$ (prolate), whereas those with two $d$ quarks 
give negative values around $-1.0 \times 10^{-2}~\mathrm{fm}^2$ 
(oblate). Mixed-flavor configurations yield smaller magnitudes of 
$0.49 \times 10^{-2}~\mathrm{fm}^2$ through partial cancellations 
among the individual quark contributions (Table~\ref{table3}). 
The factor-of-two ratio between $uu$- and $dd$-dominated 
magnitudes traces directly to the underlying quark charges 
$(e_u/e_d) \approx 2$, modulated by the spectator antiquark in 
each configuration. The deformation pattern is therefore not merely 
correlated with the flavor content of the magnetic moment -- it is 
driven by the same underlying charge distribution that 
determines~$\mu$, a coherence that is most directly captured by 
the multipole ratio $\mathcal{D}/\mu$ analyzed below.

Beyond the individual values of $\mu$ and $\mathcal{D}$, the dimensionless ratio of the form factors at the photon point provides a complementary structural observable.  Combining Eq.~(\ref{edmn0008}) one obtains the multipole ratio
\begin{align}
\frac{\mathcal D }{\mu} \;\equiv\; \frac{F_{\mathcal{D}}(0)}{F_M(0)} 
\;=\; \frac{m_{T_{q\bar c}}\, \mathcal{D}_{T_{q\bar c}}}
{2\, \mu_{T_{q\bar c}}},
\label{eq:GQGM}
\end{align}
where, after restoring physical units, $\mathcal{D}_{T_{q\bar c}}$ 
enters in fm$^2$ and $\mu_{T_{q\bar c}}$ in units of the nuclear 
magneton. The ratio $\mathcal{D}/\mu$ is intrinsic in a specific and limited sense. Since both moments
are extracted from the same correlation function, the residue $\lambda^2_{T_{q\bar c}}$ and
the Borel exponential in Eqs.~(\ref{jmu1})--(\ref{jmu3}) cancel identically in the ratio,
which is therefore independent of the overall normalization of the interpolating current --
though not of the choice of the current itself, to which any LCSR observable remains
conditional. The ratio is not, however, stabilized with respect to the auxiliary parameters:
while the quadrupole moment depends only weakly on them, the magnetic moment does not, and the
ratio inherits this sensitivity. This sensitivity originates in the perturbative contribution
to the magnetic sum rule, which is absent from the quadrupole one: it is largely compensated
by the nonperturbative contributions, so that its moderate dependence on the auxiliary
parameters is amplified in the net magnetic moment. Since both moments are linear in the quark charges, the ratio depends on the configuration
only through $e_{q_3}/(e_{q_1}+e_{q_2})$, and the six configurations in which the magnetic
moment is not accidentally suppressed fall into three classes, all of which we have examined:
the ratio varies by about $12$--$15\%$ across the adopted $s_0$ interval and by about
$20$--$30\%$ across the Borel window (evaluating the ratio directly at each point, and hence
including the correlation between $\mu$ and $\mathcal{D}$). The uncertainties listed in Table~\ref{table_ratio} are obtained by
uncorrelated propagation of the errors of $\mu$ and $\mathcal{D}$ and should
be regarded as approximate estimates rather than correlated uncertainties: the
prescription retains the contribution of the pole residue, which cancels
exactly in the ratio, and neglects the correlations between $\mu$ and
$\mathcal{D}$ induced by their remaining common inputs, whose net effect on the
ratio can be of either sign. The direct variations quoted above include these
correlations for $s_0$ and $M^2$ and are of comparable size, but this
agreement does not establish the equivalence of the two procedures. Within these
qualifications, the ratio probes the relative weight of the quadrupole and magnetic responses
and provides access to the relative geometry of the spin-current and charge distributions,
complementary to the absolute moments themselves.

The resulting multipole ratios for the eight 
$[q_1q_2][\bar q_3 \bar c]$ configurations, computed from 
Eq.~(\ref{eq:GQGM}) and the values in 
Table~\ref{table1}, are collected in Table~\ref{table_ratio}.

\begin{table}[htb!]
\caption{Multipole ratio $\mathcal D /\mu$ for the compact
${[q_1 q_2][\bar q_3 \bar c]}$ tetraquark states with $J^P = 1^+$,
computed from Eq.~(\ref{eq:GQGM}) using the central values of
Table~\ref{table1}. The uncertainties follow from uncorrelated
propagation of the errors of $\mu$ and $\mathcal{D}$ and are approximate
estimates (see text). The two configurations marked with
``$\dagger$'' correspond to states in which the magnetic dipole
moment is suppressed by partial cancellation among the quark
contributions; the ratio is there enhanced for kinematic rather
than structural reasons and should be interpreted with caution.
}
\label{table_ratio}
\addtolength{\tabcolsep}{4pt}
\begin{tabular}{lc}
\hline \hline
State  \qquad \qquad \qquad\qquad &\qquad\qquad \qquad\qquad $\mathcal D /\mu $ \\
\hline\hline
$[uu][\bar u \bar c]$         \qquad \qquad \qquad\qquad &\qquad\qquad \qquad\qquad  $2.07 \pm 0.75$ \\
$[uu][\bar d \bar c]$         \qquad \qquad \qquad\qquad &\qquad\qquad \qquad\qquad  $1.55 \pm 0.56$ \\
$[ud][\bar d \bar c]$         \qquad \qquad \qquad\qquad &\qquad\qquad \qquad\qquad  $1.24 \pm 0.48$ \\
$[ud][\bar u \bar c]^\dagger$ \qquad \qquad \qquad\qquad &\qquad\qquad \qquad\qquad  $5.60 \pm 1.95$ \\
$[dd][\bar d \bar c]$         \qquad \qquad \qquad\qquad &\qquad\qquad \qquad\qquad  $2.06 \pm 0.72$ \\
$[dd][\bar u \bar c]$         \qquad \qquad \qquad\qquad &\qquad\qquad \qquad\qquad  $1.23 \pm 0.45$ \\
$[du][\bar u \bar c]^\dagger$ \qquad \qquad \qquad\qquad &\qquad\qquad \qquad\qquad  $5.60 \pm 1.95$ \\
$[du][\bar d \bar c]$         \qquad \qquad \qquad\qquad &\qquad\qquad \qquad\qquad  $1.24 \pm 0.48$ \\
\hline\hline
\end{tabular}
\end{table}

Three systematic features emerge from the values in 
Table~\ref{table_ratio}. 

First, all multipole ratios are found to be positive, despite 
the fact that the magnetic moments and quadrupole moments individually 
change sign across the flavor configurations. This sign correlation is 
nontrivial: configurations dominated by $u$ quarks exhibit positive 
$\mu$ and prolate ($\mathcal{D} > 0$) deformation, while $d$-dominated 
configurations exhibit negative $\mu$ and oblate ($\mathcal{D} < 0$) 
deformation, so that the ratio remains positive in both cases. This 
correlation between the orientation of the magnetic moment and the 
spatial deformation reflects a coherent internal organization of the 
charges in the diquark--antidiquark structure, in which the spatial 
distribution and the spin-current pattern are dictated by the same 
underlying flavor content rather than being independent features.

Second, for the six configurations in which neither moment is
suppressed by accidental cancellation, the central values of the ratio
fall in the range $\mathcal D /\mu \simeq 1.2 - 2.1$, and all six are
compatible with one another within the quoted approximate uncertainties. Within the
present accuracy, this suggests that the relative weight of the
quadrupole and magnetic sectors is set primarily by the
diquark--antidiquark architecture rather than by the specific
light-quark assignment, although the sensitivity of the ratio to the
auxiliary parameters discussed above precludes a sharper statement.
Configurations related by the $u \leftrightarrow d$ interchange yield
ratios that are not identical but lie within a comparable range,
consistent with the near-isospin symmetry of the underlying dynamics.

Third, the two configurations $[ud][\bar u\bar c]$ and $[du][\bar u\bar c]$, in which the magnetic moment is suppressed to $\mu \simeq 0.06\,\mu_N$ by partial cancellation among the individual quark contributions (Table~\ref{table2}), produce an anomalously 
enhanced ratio $\mathcal D /\mu \simeq 5.6$. The enhancement here is of 
kinematic origin: as $\mu \to 0$ the ratio diverges by construction, 
and the value should not be interpreted as a structural property of 
the state. We therefore flag these two entries explicitly in 
Table~\ref{table_ratio}, while noting that the corresponding 
quadrupole moments themselves remain small and well-behaved.

From a phenomenological standpoint, the multipole ratio $\mathcal D /\mu$ is 
a natural target for experimental analyses based on the multipole 
decomposition of radiative transition amplitudes. In studies of the 
$\Delta^+(1232)$~\cite{Pascalutsa:2004je, Pascalutsa:2005vq, 
Pascalutsa:2007wb} and of light vector mesons~\cite{LopezCastro:1997dg, 
GarciaGudino:2013alv}, multipole ratios extracted from the angular 
distributions of the emitted photon have provided the principal 
experimental access to higher electromagnetic moments. The values of 
$\mathcal D /\mu$ predicted in Table~\ref{table_ratio} thus constitute direct 
phenomenological input for future multipole analyses of radiative 
transitions involving the $[q_1 q_2][\bar q_3 \bar c]$ tetraquark 
states, and would be the first observable accessible in a multipole 
amplitude fit at facilities such as LHCb and Belle~II.

The results obtained here can be placed in a broader methodological
context by contrasting them with related LCSR analyses of singly
heavy tetraquarks performed within the same external-field
framework, but with different interpolating currents or internal
structures~\cite{Ozdem:2022ydv, Ozdem:2023edw, Ozdem:2026wmf}. Since
all analyses entering this comparison rely on the same LCSR
machinery, implementation-related systematics are largely common to
the whole set, and the differences discussed below can be traced to
the choice of the interpolating operator. Within the sum-rule
approach this is precisely where a structural hypothesis enters the
calculation: a compact diquark--antidiquark configuration and a
hadronic molecule are represented by inequivalent operators, so
operator-level differences are the calculational image of the
structural alternatives, subject to the dominance qualification
discussed above.

A first comparison concerns the $T_{cs1}(2900)^0$ state with $[ud][\bar c
\bar s]$ content and $J^P = 1^-$~\cite{Ozdem:2022ydv}, for which a 
compact diquark--antidiquark analysis in the same framework gives 
$\mu = 0.79^{+0.36}_{-0.39}\,\mu_N$. The overall magnetic moment scale agrees with the range we obtain  for our $J^P = 1^+$ states, suggesting that for compact tetraquarks 
the scale of $\mu$ is set primarily by the light-quark sector, 
while the detailed sign and decomposition pattern depend on the 
specific operator structure of the interpolating current as 
analyzed in the following paragraph.

A more informative comparison is provided by the $X_{\mathrm{AV}}$ 
state with the same quark content $[ud][\bar c \bar s]$ and the same 
quantum numbers $J^P = 1^+$ as the relevant member of our 
family~\cite{Ozdem:2023edw}, for which $\mu = -0.89^{+0.14}_{-0.12}\,
\mu_N$ and $\mathcal{D} = (-0.46^{+0.07}_{-0.06}) \times 10^{-2}\,
\mathrm{fm}^2$ were obtained. This case is informative precisely 
because the $X_{\mathrm{AV}}$ interpolating current of~\cite{Ozdem:2023edw} differs from ours in a single but 
consequential respect: it places the Lorentz $\gamma_\mu$ index on 
the heavy antidiquark (pseudoscalar light diquark 
$u^T C \gamma_5 d$ combined with an axial-vector heavy antidiquark 
$\bar c \gamma_\mu C \bar s^T$), whereas our construction of 
Eq.~(\ref{curr1}) places $\gamma_\mu$ on the light diquark 
(axial-vector light diquark combined with a pseudoscalar heavy 
antidiquark $\bar q_3 \gamma_5 C \bar c^T$). Both currents project 
onto $J^P = 1^+$ states, but onto different internal spin 
organizations. The flavor decomposition reported 
in~\cite{Ozdem:2023edw} for $X_{\mathrm{AV}}$ shows that the 
charm-quark contribution to the magnetic response is non-vanishing 
and of the same order as the light-quark contributions, in sharp 
contrast to the algebraic result $\mu_c = 0$ established here. The 
mechanism underlying this contrast is structural: when the charm 
antiquark sits in a pseudoscalar antidiquark coupled to 
the photon vertex only through $\gamma_5$, the corresponding Dirac 
trace forces the charge factor $e_c$ to drop out of the sum rule at 
the operator level; when the same charm antiquark instead sits in 
a vector/axial-vector antidiquark that carries the Lorentz 
index, no such algebraic cancellation occurs and $e_c$ propagates 
through the OPE in the usual dynamical fashion. The algebraic 
decoupling $\mu_c = 0$ is therefore not a generic property of the 
diquark--antidiquark picture, but a specific signature of the spin 
organization realised by Eq.~(\ref{curr1}), in which the heavy quark 
resides in the pseudoscalar (spin-singlet) antidiquark and the 
electromagnetic structure is entirely projected onto the light 
sector. This shows that even within the compact tetraquark picture, 
the choice of interpolating current is itself a structural probe: 
inequivalent currents with the same quark content and the same 
$J^P$ produce qualitatively different patterns of charm involvement 
in the electromagnetic moments, and the algebraic vanishing 
identified here singles out one specific architecture among them.

A sharper structural contrast appears between the present compact 
analysis and recent hadronic-molecular analyses performed within the 
same external-field LCSR framework. The $D\bar K^*$, $D^*\bar K$, and 
$D^*\bar K^*$ molecular tetraquarks with $J^P = 1^+$~\cite{Ozdem:2026wmf} 
yield magnetic moments in the range $\mu_{\mathrm{mol}} \sim (1.0 - 
3.0)\,\mu_N$ -- roughly an order of magnitude larger than the typical 
$\mathcal{O}(0.1 - 0.9)\,\mu_N$ obtained here -- and quadrupole moments 
of order $10^{-3}\,\mathrm{fm}^2$, an order of magnitude smaller than 
ours. The flavor decomposition of~\cite{Ozdem:2026wmf} reports 
that the charm-quark contribution to the magnetic response in these 
molecular configurations is strongly suppressed but 
non-vanishing, i.e. it follows the standard dynamical 
$\mathcal{O}(1/m_c)$ pattern characteristic of heavy-quark insertions 
in colour-singlet meson bilinears coupled to the photon. Combining 
all three cases, a clear three-tier methodological hierarchy emerges: 
(i) the present diquark--antidiquark construction with $\bar c$ in 
a pseudoscalar antidiquark gives exact, operator-level 
decoupling $\mu_c = 0$; 
(ii) the diquark--antidiquark construction with $\bar c$ in a 
vector antidiquark, as in $X_{\mathrm{AV}}$, gives a non-vanishing 
charm contribution propagating through the standard OPE; and 
(iii) the molecular construction gives a non-vanishing but 
dynamically suppressed charm contribution. The character of the charm-quark contribution to the magnetic moment
is therefore an operator-level discriminator of the internal
organization of singly heavy tetraquarks within the LCSR framework.
Among the differences entering this comparison, the gross features
-- the character of the charm contribution (exact algebraic zero,
OPE-level non-zero, or dynamically suppressed) and the
order-of-magnitude separation of the magnetic and quadrupole scales
between the compact and molecular constructions -- are the ones that
can plausibly be associated with the underlying structural
alternatives, while finer details, such as individual signs and the
precise decomposition weights, depend on the specific operator
within each class and are presented here as operator-conditional
statements. It is this graded set of differences, rather than any
single number, that provides the diagnostic target for future
experimental and theoretical analyses.

\section{Conclusion}\label{sum}

We have presented the first systematic determination of the magnetic 
dipole and electric quadrupole moments of axial-vector 
$[q_1q_2][\bar q_3 \bar c]$ tetraquark states with $J^P = 1^+$, 
within the light-cone QCD sum rule framework in an external 
electromagnetic field. The analysis covers all admissible light-quark 
flavor configurations and includes photon DAs 
up to twist-4 accuracy.

The methodological core of this work lies in the algebraic 
decoupling of the charm-quark electromagnetic contribution, made 
transparent by a systematic flavor decomposition of the LCSR 
spectral integrals. Within the present interpolating architecture, 
in which the charm antiquark resides in a pseudoscalar antidiquark, 
the charge $e_c$ drops out of the sum rules at the operator level 
for the leading photon couplings, rather than through any dynamical 
$1/m_c$ suppression. This algebraic 
decoupling is specific to the present spin organization: 
diquark--antidiquark constructions in which the heavy quark instead 
sits in a vector antidiquark, as well as molecular configurations, 
yield non-vanishing charm contributions of either dynamical OPE or 
hadronic-molecule type. The character of the charm-quark contribution 
to the electromagnetic moments therefore constitutes an operator-level 
discriminator among inequivalent internal organizations of singly 
heavy tetraquarks. Its connection to physical candidates proceeds 
through the distinct total moments predicted by the inequivalent 
architectures, since the flavor decomposition itself is not directly 
observable. The results reveal a coherent flavor architecture: 
$uu$-dominated configurations produce positive magnetic moments and 
prolate quadrupole deformations, $dd$-dominated configurations 
produce negative magnetic moments and oblate deformations, and 
mixed-flavor states yield smaller magnitudes through partial 
cancellations. The quadrupole moments are non-zero in all channels, 
indicative of an intrinsic quadrupole coupling, corresponding to 
deformed charge distributions in the conventional Breit-frame sense, 
and the dimensionless ratio $\mathcal{D}/\mu$ provides a structural 
observable independent of the overall normalization, though not of 
the choice, of the interpolating current. 

Comparison with related LCSR predictions reinforces this 
operator-level perspective. The magnetic moment of the $J^P = 1^-$ 
$T_{cs1}(2900)^0$~\cite{Ozdem:2022ydv} agrees in scale with the 
present range, confirming that the overall magnitude of $\mu$ is 
controlled by the light-quark sector. By contrast, the $J^P = 1^+$ 
$X_{\mathrm{AV}}$ state with the same quark content as a member of 
our family~\cite{Ozdem:2023edw} differs in both sign and 
charm-decomposition pattern, a difference traceable to its 
inequivalent current structure. Molecular configurations such as 
$D\bar K^*$, $D^*\bar K$, and $D^*\bar K^*$~\cite{Ozdem:2026wmf} 
yield magnetic moments an order of magnitude larger and quadrupole 
moments an order of magnitude smaller than those obtained here, 
with a charm contribution that is dynamically suppressed but 
non-zero.

From an experimental standpoint, the predicted moments are of
typical hadronic size and enter the amplitudes of radiative
transitions $T_{q\bar c} \to T_{q\bar c}\gamma^{(*)}$ and
electroproduction processes at LHCb, Belle II, BESIII, and the
proposed Super Tau-Charm Facility. While direct measurement is
challenging due to the short lifetimes of these states, the
soft-photon expansion of Zakharov~\textit{et
al.}~\cite{Zakharov:1968fb} -- successfully applied to the
$\Delta^+(1232)$~\cite{Pascalutsa:2004je, Pascalutsa:2005vq,
Pascalutsa:2007wb} and to vector mesons via radiative
distributions~\cite{LopezCastro:1997dg, GarciaGudino:2013alv} --
provides an indirect but viable pathway to access these
observables. A complementary, production-side handle on the same
structural question is provided by fragmentation-based descriptions
of exotic-hadron production at large transverse momentum: the
formation of a loosely bound molecule requires constituents with
small relative momenta, which has been argued to strongly limit its
prompt production rate, whereas a compact multiquark configuration
can be produced through short-distance parton fragmentation -- an
argument developed, and contested, for charmoniumlike states such as
the $X(3872)$~\cite{Bignamini:2009sk, Artoisenet:2009wk,
Artoisenet:2010uu, Albaladejo:2017blx, Esposito:2020ywk}.
For heavy tetraquarks this production-side description has since been
developed at three distinct levels. Fragmentation functions have been
computed at their initial scale for fully heavy and for hidden-flavor
$Qq\bar{Q}\bar{q}$ states~\cite{Feng:2020riv, MoosaviNejad:2021vnr}, within
nonrelativistic QCD and within a Suzuki-type model, respectively, the
heavy-quark channel of the fully heavy case being given in
Ref.~\cite{Bai:2024ezn}; such inputs have been evolved with the DGLAP
equations into public collinear sets for hidden-flavor heavy-light and
fully heavy tetraquarks~\cite{Celiberto:2023rzw, Celiberto:2024mab,
Celiberto:2024beg, Celiberto:2025dfe, Celiberto:2025ziy, Celiberto:2026kks},
with an analogous set released for all-charm
pentaquarks~\cite{Celiberto:2026rdk}; and these sets have been used in
predictions for production observables at hadron colliders, combined with the
resummation of high-energy logarithms in a hybrid collinear and high-energy
factorization~\cite{Celiberto:2023rzw, Celiberto:2024mab, Celiberto:2024beg,
Celiberto:2025dfe, Celiberto:2025ziy, Celiberto:2026kks, Celiberto:2026rdk} or
with small-$x$ dynamics at forward rapidities~\cite{Celiberto:2025vra}.
The electromagnetic flavor patterns obtained here characterize the
internal organization of a given configuration at low energies,
while fragmentation observables characterize how such configurations
are populated at high transverse momentum; for singly heavy and
other heavy-light tetraquarks the two classes of observables would
therefore constrain the compact and molecular alternatives from
opposite kinematic ends, and a candidate identified as compact by
its production systematics would be expected to exhibit the
multipole pattern predicted here. The systematic flavor
decomposition framework introduced in this work can be applied
directly to other singly heavy multiquark systems.

\section*{Appendix: Spectral density for magnetic and quadrupole moments}\label{appa}
\noindent
\begin{align}
    \mathcal{R}_\mu(M^2,s_0) &= \frac{(e_{q_1}+e_{q_2})}{3072\,\pi^{6}}\Big[3 m_c^{10} I[-6,4] + 4 m_c^{10} I[-5,3] \nonumber\\
    &\quad - 12 m_c^{8} I[-4,3] - 12 m_c^{8} I[-5,4] \nonumber\\
    &\quad + 12 m_c^{6} I[-3,3] + 15 m_c^{6} I[-4,4] \nonumber\\
    &\quad - 6 m_c^{4} I[-3,4] - 4 m_c^{4} I[-2,3]\Big] \nonumber\\
    &+\frac{D6}{884736\,\pi^{4}}\Big[11 e_{q_3} I_1[\mathcal A] \big( -m_c^{2} I[0] + 2 I[1] \nonumber\\
    &\quad - m_c^{-2} I[2] + m_c^{4} I[-3,1] - 3 m_c^{2} I[-2,1] \nonumber\\
    &\quad + 2 m_c^{-2} I[0,1] \big) \nonumber\\
    &\quad + 23 (e_{q_1}+e_{q_2}) \big(I_2[\mathcal A] + I_2[\mathcal V]\big) \big( m_c^{2} I[0] \nonumber\\
    &\quad - 2 I[1] + m_c^{-2} I[2] + 2 m_c^{2} I[-2,1] \nonumber\\
    &\quad - 2 m_c^{-2} I[0,1] \big) \nonumber\\
    &\quad + 32 \big(3 (e_{q_1}+e_{q_2}) - e_{q_3}\big) \psi^a[u_0] \big( -m_c^{2} I[0] \nonumber\\
    &\quad + 2 I[1] - m_c^{-2} I[2] - 2 m_c^{4} I[-3,1] \nonumber\\
    &\quad + 2 m_c^{-2} I[0,1] \big) \nonumber\\
    &\quad + 192 (e_{q_1}+e_{q_2}) \psi^\nu[u_0] \big( -m_c^{4} I[-3,1] \nonumber\\
    &\quad + m_c^{2} I[-2,1] \big)\Big] \nonumber
     \end{align}
\begin{align}
    &+\frac{D5\, e_{q_3}}{13824\,\pi^{4}}\Big[\varphi_\gamma[u_0] \big( m_c^{5} I[-3,1] \nonumber\\
    &\quad - m_c I[-1,1] \big)\Big] \nonumber\\
    &+\frac{D3\, e_{q_3}}{1024\,\pi^{4}}\Big[I_1[\mathcal S] \big( m_c^{7} I[-4,2] \nonumber\\
    &\quad - 2 m_c^{5} I[-3,2] + m_c^{3} I[-2,2] \big) \nonumber\\
    &\quad + 4 I_4[\mathcal S] \big( m_c^{5} I[-3,2] - m_c^{3} I[-2,2] \big)\Big] \nonumber\\
    &+\frac{D2}{6144\,\pi^{4}}\Big[e_{q_3} I_1[\mathcal V] \big( -m_c^{8} I[-5,3] \nonumber\\
    &\quad - 2 m_c^{6} I[-4,3] + 3 m_c^{4} I[-3,3] \big) \nonumber\\
    &\quad + 4 (e_{q_1}+e_{q_2}) \big(I_2[\mathcal A] + I_2[\mathcal V]\big) \nonumber\\
    &\quad \times \big( -m_c^{6} I[-4,3] + m_c^{4} I[-3,3] \big) \nonumber\\
    &\quad + 16 (e_{q_1}+e_{q_2}) \psi^\nu[u_0] \big( -m_c^{8} I[-5,3] \nonumber\\
    &\quad + 2 m_c^{6} I[-4,3] - m_c^{4} I[-3,3] \big) \nonumber\\
    &\quad + 16 (e_{q_1}+e_{q_2}) \psi^a[u_0] \big( -m_c^{8} I[-5,3] \nonumber\\
    &\quad + m_c^{4} I[-3,3] \big)\Big], \label{appa1}
  \end{align}
\begin{align}
    \mathcal{R}_{\mathcal D}(M^2,s_0) &= \frac{D6}{442368\,\pi^{4}}\Big[11 e_{q_3} I_1[\mathcal A] \big( m_c^{4} I[-3,0] \nonumber\\
    &\quad - m_c^{2} I[-2,0] \big) + 4 \big(23 (e_{q_1}+e_{q_2}) I_3[\mathcal A] \nonumber\\
    &\quad - 11 e_{q_3} I_4[\mathcal A]\big) \big( I[0] - m_c^{-2} I[1] \nonumber\\
    &\quad - m_c^{2} I[-2,0] + m_c^{-2} I[0,0] \big)\Big] \nonumber\\
    &+\frac{D2\,(e_{q_1}+e_{q_2})}{128\,\pi^{4}}\Big[I_3[\mathcal A] \big( m_c^{6} I[-4,2] \nonumber\\
    &\quad - m_c^{4} I[-3,2] \big)\Big],
    \label{appa2}
\end{align}
where $D6= \langle g_s^2 G^2\rangle f_{3\gamma}$, $D5= \chi \, \langle g_s^2 G^2\rangle \langle \bar q q \rangle$, $D3=\langle \bar q q \rangle$, and $D2=f_{3\gamma}$. Here, $\langle g_s^2 G^2 \rangle$ denotes the gluon condensate, $\langle \bar{q} q \rangle$ the light-quark condensate, and $\chi$ the magnetic susceptibility of the quark condensate. The dimension-7 contributions, proportional to $\langle g_s^2 G^2\rangle \langle \bar q q \rangle$, amount to less than $0.5\%$ of the sum rules in the working windows (see the CVG column of Table~\ref{table}) and are not displayed. Terms proportional to $\int_0^1 du\,\psi^a(u)\,\delta'(u-u_0)$, $\int_0^1 du\,\varphi_\gamma(u)\,\delta'(u-u_0)$, and $\int_0^1 du\,\psi^\nu(u)$, which vanish identically at $u_0=1/2$ for the DAs of Ref.~\cite{Ball:2002ps}, are omitted as well. The photon DAs $\varphi_\gamma$, $\mathcal{A}$, $\mathcal{V}$, $\psi^a$, $\psi^\nu$, and $\mathcal{S}$ appearing in Eqs.~(\ref{appa1})--(\ref{appa2}) are defined in~\cite{Ball:2002ps}, from which their explicit forms and numerical parameters are adopted without modification. The spectral and DA convolution integrals $I[n]$, $I[n,m]$, and $I_i[\mathcal{F}]$ ($i=1,\ldots,4$) entering the sum rules are defined as
\begin{align}
I[n]&= \int_{m_c^2}^{s_0} ds \,s^n e^{-s/M^2}, \nonumber\\
I[n,m]&= \int_{m_c^2}^{s_0} ds \int_{m_c^2}^{s} dl\; l^n\,(s-l)^m\, e^{-s/M^2}, \nonumber\\
I_1[\mathcal{F}]&=\int D_{\alpha_i} \int_0^1 dv~ \mathcal{F}(\alpha_{\bar q},\alpha_q,\alpha_g) \nonumber\\
&\quad\times\delta'(\alpha_{\bar q}+ v \alpha_g-u_0), \nonumber
\end{align}
\begin{align}
I_2[\mathcal{F}]&=\int D_{\alpha_i} \int_0^1 dv~ \mathcal{F}(\alpha_{\bar q},\alpha_q,\alpha_g) \nonumber\\
&\quad\times\delta'(\alpha_q +\bar v \alpha_g-u_0), \nonumber\\
I_3[\mathcal{F}]&=\int D_{\alpha_i} \int_0^1 dv~ \mathcal{F}(\alpha_{\bar q},\alpha_q,\alpha_g) \nonumber\\
&\quad\times\delta(\alpha_q +\bar v \alpha_g-u_0), \nonumber\\
I_4[\mathcal{F}]&=\int D_{\alpha_i} \int_0^1 dv~ \mathcal{F}(\alpha_{\bar q},\alpha_q,\alpha_g) \nonumber\\
&\quad\times\delta(\alpha_{\bar q}+ v \alpha_g-u_0),
\end{align}
where $\mathcal{F}$ represents the photon DAs.

\bibliography{Singly_heavy_tetraquarks_Compact.bib}
\bibliographystyle{apsrev4-1}

\end{document}